\documentclass[preprint,aps]{revtex4-1}
\usepackage{graphicx}
\usepackage{textcomp}
\usepackage{color}
\begin{document}

\title{Critical current anisotropy in Nd-1111 single crystals and the influence of neutron irradiation}
\author{M.~Eisterer}
\author{V.~Mishev}
\author{M.~Zehetmayer}
\address{Atominstitut, Vienna University of Technology, Stadionallee 2, 1020 Vienna, Austria}
\author{N.~D.~Zhigadlo}
\address{Laboratory for Solid State Physics, ETH Zurich, CH-8093 Zurich, Switzerland}
\author{S. Katrych}
\author{J.~Karpinski}
\address{Laboratory for Solid State Physics, ETH Zurich, CH-8093 Zurich, Switzerland}
\address{Institute of Condensed Matter Physics, EPFL, CH-1015 Lausanne, Switzerland}

\begin{abstract}
We report on angle-resolved magnetization measurements on NdFeAsO$_{0.65}$F$_{0.35}$ (Nd-1111)
single crystals. The field dependence of the critical current
density, $J_\mathrm{c}$, is non-monotonous in these crystals at all orientations and
temperatures due to the fishtail effect, which strongly influences the
angular dependence of $J_\mathrm{c}$. The currents decrease as the
field is tilted from the crystallographic $c$-axis at low fields, but increase at high fields. A peak occurs in the angular
dependence of $J_\mathrm{c}$ at intermediate
fields. The critical currents are significantly enhanced after
irradiation with fast neutrons and the fishtail disappears. The
different current anisotropies at low and high fields, however,
persist. We discuss the data in the framework of the anisotropic
scaling approach and propose a transition from dominant pinning by
large defects of low density at low fields to pinning by small
defects of high density at high fields in the pristine crystal.
Strong pinning dominates at all fields after the irradiation, and
the angular dependence of $J_\mathrm{c}$ can be described by
anisotropic scaling only after an appropriate extension to this pinning regime.
\end{abstract}

\maketitle
\section{Introduction}
The pinning properties of iron-based superconductors have been in the focus of intensive research since the discovery of superconductivity in these compounds.\cite{Kam08} Many similarities to the cuprates were found. The critical currents in single crystals often show a fishtail effect, which disappears after the introduction of an efficient pinning landscape, for instance by irradiation techniques.\cite{Eis09c,Eis09b,Pro10,Fan12,Tam12,Sha13}
In thin films on the other hand, pinning is much stronger and the currents decrease monotonously with field.\cite{Lee10,Hae11,Iid11b,Mai11,Hab12,Mai12,Tar12,Miu13,Eis11,Bel12,Mel12,Che12,Iid13,Kid11,Tak12,Iid13b}
Angle-resolved measurements of the pinning properties are very efficient for studying the pinning landscape and anisotropy effects of the vortex lattice. They were performed nearly exclusively on films so far, in which growth-related and often correlated defects dominate the properties, which are not representative for the defects prevailing in bulk materials such as single crystals or grains in wires or tapes. Angle-resoved measurements on crystals are thus highly desirable to complement the film data.
Thin films are widely available only of the Ba-122 (BaFe$_2$As$_2$) \cite{Lee10,Hae11,Iid11b,Mai11,Hab12,Mai12,Tar12,Miu13} and 11 (FeSe$_{1-x}$Te$_x$) \cite{Eis11,Bel12,Mel12,Che12,Iid13} families and only a few data exist for the 1111 (LaFeAsO) family.\cite{Kid11,Tak12,Iid13b} The latter has the highest anisotropy among them, \cite{Put10} which makes it the best candidate for studying anisotropy effects. Anisotropy is considered as a key parameter for applications, since it enhances the harmful thermal fluctuations.
In this study we report on the anisotropy of the in-plane critical currents of Nd-1111 single crystals by angle-resolved magnetization measurements. The results are discussed in the framework of the anisotropic scaling approach.\cite{Bla92} After the characterization of the pristine crystals the defect structure was changed completely by irradiation with fast neutrons to assess changes in the pinning properties arising from the introduced pinning centers.

\section{Experimental\label{secexp}}

The NdFeAsO$_{0.65}$F$_{0.35}$ single crystals were prepared by a high pressure technique.\cite{Zhi12} Two crystals were studied, whose geometries were determined in two steps. First, an optical microscope was used to establish the lateral surface area. A subsequent mass measurement enabled the calculation of the volume and the thickness of the samples from the theoretical mass density.\cite{Kar09b} The results are listed in Table ~\ref{tabGeo}. The transition temperature ($T_\mathrm{c}$) was measured in a 1 T SQUID by applying an AC field of 0.3 mT. The reported transition temperature refers to the onset of superconductivity, where the susceptibility starts to deviate from its behavior in the normal conducting state.

\begin{table}[h]
\begin{center}
    \begin{tabular}{|c|c|c|c|}
    \hline
    Sample & a (mm) & b (mm) & c (mm) \\ \hline
    Nd1111\#1 & 0.633 & 0.401 & 0.058 \\ \hline
    Nd1111\#2 & 0.497 & 0.36 & 0.0309 \\ \hline
    \end{tabular}
\caption{Sample geometries}
\label{tabGeo}
\end{center}
\end{table}
Magnetization loops were recorded on crystal \#1 at different
temperatures in a 7 T SQUID with the field applied parallel
to the $c$-axis of the sample. The critical current density,
$J_\mathrm{c}$, along the $ab$-planes was evaluated from the
irreversible magnetic moment $m_\mathrm{irr}$. A self-field
correction was applied for the calculation of the average magnetic field $B$ within the crystal.\cite{Zeh09} 

Sample \#1 and \#2 were irradiated to a fast neutron fluence ($E>0.1$\,MeV) of $3.7 \cdot
10^{21}$\,m$^{-2}$ and $1.8 \cdot
10^{21}$\,m$^{-2}$, respectively. The fluence was determined from the
radioactivity of a nickel foil which was placed in the same quartz
tube as the sample during the irradiation. Fast neutron irradiation is
known to result in a variety of defects, ranging from single
displaced atoms to spherical defect cascades of about 5\,nm in
diameter.\cite{Fri93,Ale98,Eis09b,Chu12} 

Angle-resolved magnetization measurements
were performed on crystal \#2 in a 5 T vector Vibrating Sample
Magnetometer (VSM). Previous studies described similar measurements
on superconducting thin films,\cite{Tho10,Hen11, Eis11} where
the currents flow parallel to the lateral surface at all orientations. The
measurements on crystals can be interpreted in the same way as
long as the currents remain parallel to the $ab$-planes (the large surface).  This
condition was verified from the orientation of the magnetic moment,
which is available in a vector VSM. The currents inside the sample
were found to remain parallel to the large surface (and to the $ab$-planes) up
to an angle of at least 80\textdegree, which is a consequence of the large
aspect ratio of the crystal. Only data within this angular range will be
considered in the following in order to avoid problems with
currents flowing in arbitrary directions and the resulting change
in geometry of the current loops. However, not all currents
flow under Maximum Lorentz Force (MLF), when the sample is
inclined from one of its main orientations, and the currents
flowing under Variable Lorentz Force (VLF) potentially change the
angular dependence of $J_\mathrm{c}$.\cite{Eis11} Whenever the VLF-currents may influence the behavior in a qualitative way, it will be noted explicitly. We will also restrict our
considerations to 15\,K where the VSM signal is sufficiently large ($>
10^{-7}$\,Am$^{2}$), the self-field is comparatively small and
the peak of the fishtail is visible in a wide angular range. At
higher temperatures, the signal of the tiny crystals was too small
for a careful analysis, at lower temperatures the self field
increases and the fishtail moves out of the accessible field range ($<5$\,T) at
rather low angles. However, the behavior did not change qualitatively at these temperatures.


\section{Results \label{SecRes}}

\begin{figure} \centering \includegraphics[clip,width=0.9\textwidth]{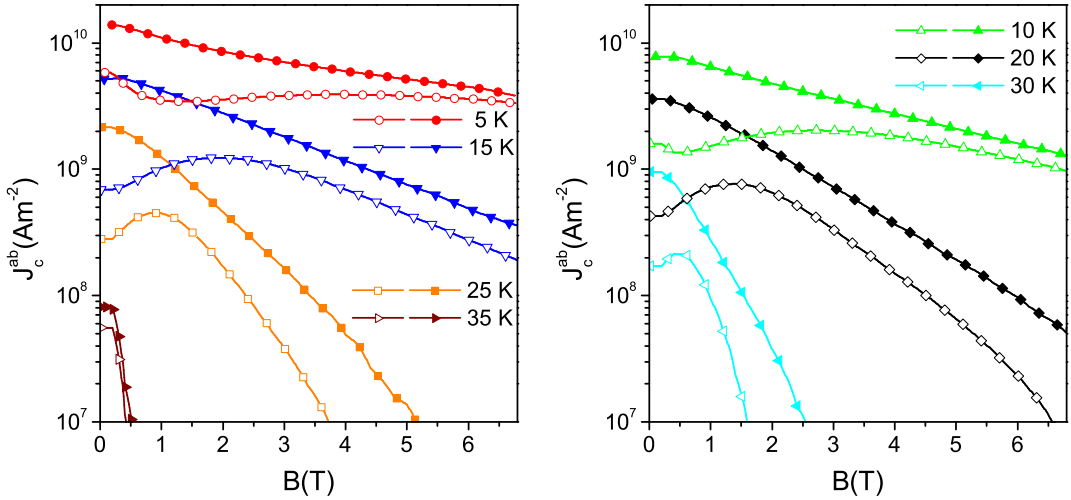}
\caption{Critical current densities of the Nd-1111 single crystal \#1 at various temperatures as a function of the magnetic flux density ($B\|c$). The open and solid symbols refer to the pristine and the neutron irradiated crystal, respectively.} \label{FigJcc}
\end{figure}

The irradiation slightly reduces the transition temperature from 39.9\,K to 39.3\,K in crystal \#1 ($3.7\cdot 10^{21}\,m^{-2}$) and
from 39.3\,K to 39.1\,K in crystal \#2 ($1.8\cdot 10^{21}\,m^{-2}$). These findings are consistent with previous reports on Sm-1111 bulk samples \cite{Eis09b} or Ba-122 single crystals.\cite{Eis09c} The modest decrease in $T_\mathrm{c}$ is also comparable with that in the cuprates \cite{Sau98,Wei10}. A small neutron fluence does not harm the transition temperature significantly, but improves pinning. (Note that superconductivity is totally suppressed after irradiation to a neutron fluence of the order of $10^{23}\,m^{-2}$.\cite{Kar09}.)

Figure~\ref{FigJcc} shows the changes in critical current density
upon neutron irradiation at various temperatures for the magnetic
field applied parallel to the crystallographic $c$-axis. A strong
increase in $J_\mathrm{c}$, the disappearance of the fishtail (or second peak) effect and a shift
of the irreversibility field at high temperatures are observed.
This behavior resembles the corresponding changes in cuprate
superconductors,\cite{Wer00,Wis00,Wei10} Sm-1111 bulk samples,\cite{Eis09b} and
Sm-1111 crystals irradiated with heavy ions.\cite{Fan13} In the latter case,
the enhancement as well as the resulting currents are much higher, because this
irradiation technique introduces larger defects and because of the higher
transition temperature of those 1111 crystals, which were much closer
to optimal doping than the crystals of our study. In Co-doped
Ba-122 single crystals on the other hand, the irreversibilty
fields tend to decrease at high temperatures after fast neutron irradiation, while similar
$J_\mathrm{c}$-enhancements were found.\cite{Eis09c} 

\begin{figure} \centering \includegraphics[clip,width=0.9\textwidth]{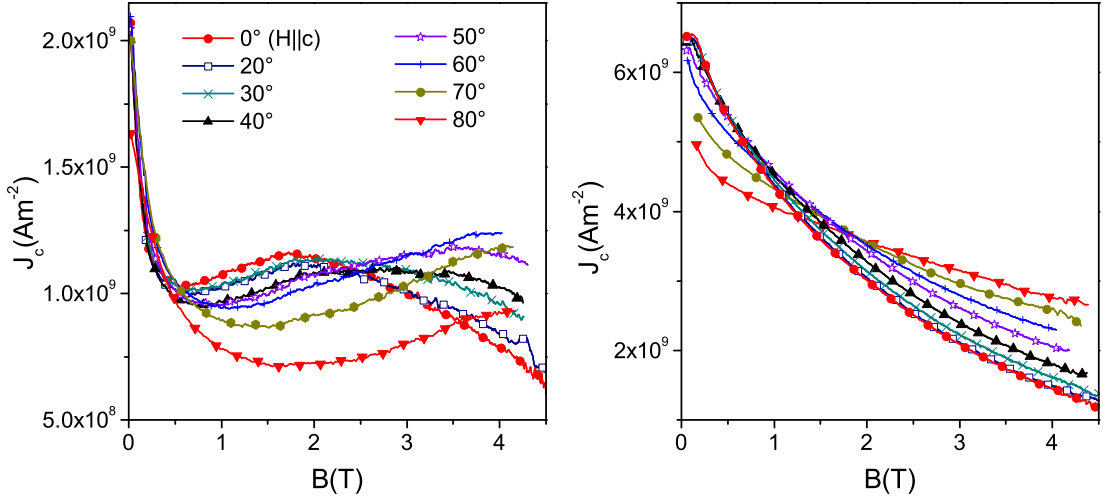}
\caption{Field dependence of the critical current density in crystal \#2 at 15\,K for various angles between the applied magnetic field and the crystallographic $c$-axis. Left panel: pristine crystal. Right panel: after irradiation to a fast neutron fluence of $1.8 \cdot 10^{21}$\,m$^{-2}$} \label{FigJc15K}
\end{figure}

\begin{figure} \centering \includegraphics[clip,width=0.9\textwidth]{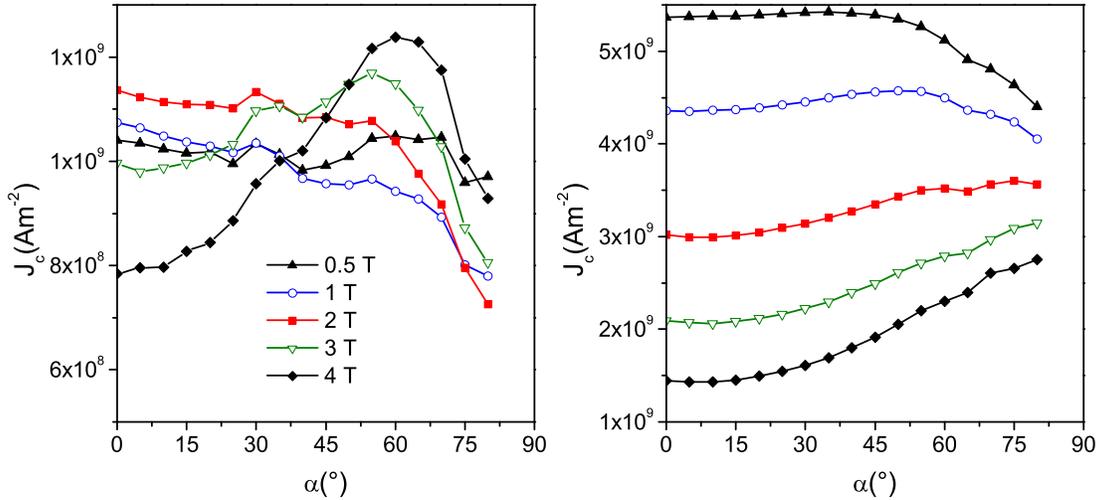}
\caption{Angular-dependence of the critical currents (crystal \#2) at 15\,K prior to (left) and after (right) irradiation.} \label{Figangular}
\end{figure}

Next, we consider the anisotropy of the critical currents including the
influence of disorder. The field dependence of $J_\mathrm{c}$
at 15\,K and varying crystal orientation is shown in
Fig.~\ref{FigJc15K}.  $\alpha$ denotes the angle between the applied
magnetic field and the crystallographic $c$-axis, thus
$\alpha=0$ refers to $H\| c$. In the left panel (pristine crystal), the position of the ``fishtail''-peak shifts to higher magnetic fields at larger $\alpha$ and the peak value of $J_\mathrm{c}$ grows for $\alpha\gtrsim 50$\textdegree. Below the
peak field, the currents decrease with $\alpha$, in contrast to
expectations for uncorrelated pinning centers in an anisotropic
superconductor. At high fields, the ``usual'' behavior,
i.e. growing currents with increasing $\alpha$, is found. The
angular dependence of $J_\mathrm{c}$ is plotted in
Fig.~\ref{Figangular} for a better illustration of the change in
behaviour. A peak occurs in $J_\mathrm{c}(\alpha)$ at 3 and 4\,T, because these fields are above and below the position of the ``fishtail''-peak at low ($\lesssim 40$\textdegree) and high angles ($\gtrsim 65$\textdegree), respectively.

Although $J_\mathrm{c}$ becomes monotonous with field after
irradiation, a similar transition from the ``unusual'' behavior at
low fields to the expected behavior at high fields is observed (right panel in Fig.~\ref{Figangular}),
since the $J_\mathrm{c}(B)$ curves cross each other (right panel in Fig.~\ref{FigJc15K}).

\section{Discussion\label{SecDis}}

The current standard approach for modelling anisotropy effects in superconductors was proposed by Blatter et al. more than two decades ago.\cite{Bla92} The main idea of this approach consists of scaling all relevant superconducting properties by functions of $\epsilon(\alpha)$, which is given by
\begin{equation}
\epsilon(\alpha)=\sqrt{\gamma^{-2} \sin^2(\alpha)+ \cos^2(\alpha)}
\nonumber
\end{equation}
The anisotropy parameter $\gamma$ originally refers to the
anisotropy of the effective mass of the charge carriers but is
usually determined by the anisotropy of the upper critical field
(i.e. $\gamma=B^{ab}_\mathrm{c2}/B^{c}_\mathrm{c2}$, where the
indices $ab$ and $c$ refer to the crystallographic $ab$-planes and
$c$-axis, respectively). In particular, the angular dependence of
the upper critical field becomes $B_\mathrm{c2}(\alpha)=B^{c}_\mathrm{c2}/\epsilon(\alpha)$, as
predicted by anisotropic Ginzburg-Landau theory, thus
motivating the anisotropic scaling approach. This behavior is
widely observed in many classes of superconductors, although
multi-band \cite{Gur03,Ara05} or two-dimensional \cite{Mar92}
superconductivity may cause deviations. Available data on the iron
based superconductors \cite{Eis09c,Iid10} suggest its validity also in
this new family. The irreversibilty fields are expected to share the same angular
dependence
($B_\mathrm{irr}(\alpha)=B^{c}_\mathrm{irr}/\epsilon(\alpha)$), if
pinning is not too anisotropic. Since $B_\mathrm{irr}$ defines the
field where $J_\mathrm{c}$ becomes zero, it is obvious that the
angular dependence of $J_\mathrm{c}$ at high fields (close to
$B_\mathrm{irr}$) has to be dominated by the behavior of
$B_\mathrm{irr}$ itself. This is indeed observed in both the
pristine and irradiated crystal.

The scaling law for $J_\mathrm{c}$ is less obvious because
$J_\mathrm{c}$ is given by the (extrinsic) pinning properties. The
original prediction of the anisotropic scaling approach is based
on the collective pinning theory, which was proposed for a high
density of weak pinning sites. In this case and if the currents flow parallel to the $ab$-planes, only the field has to
be scaled:
$J_\mathrm{c}(B,\alpha)=J_\mathrm{c}(B\epsilon(\alpha),0)$.
This means that the field virtually decreases if the sample is
rotated from 0 to 90\textdegree. If the currents decrease with field (as usual), they increase with $\alpha$. Due to the
fishtail effect in the pristine samples of our study the currents
\emph{increase} with field in a certain field range, where the
scaling approach results in \emph{decreasing} currents at increasing
$\alpha$. However, this does not explain all features of the
angular dependence observed in the unirradiated crystal, as discussed in the following.

\begin{figure} \centering \includegraphics[clip,width=0.9\textwidth]{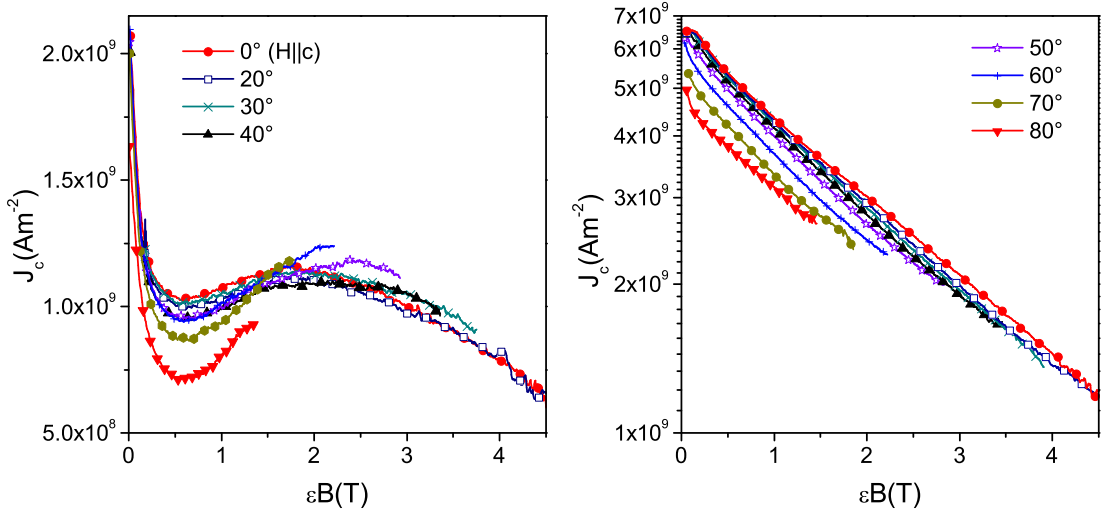}
\caption{Same data as in Fig.~\ref{FigJc15K} but with field scaling assuming an anisotropy $\gamma$ of 3.5.} \label{FigScal}
\end{figure}

The data of Fig.~\ref{FigJc15K} are replotted as a function of the
scaled field ($B\epsilon(\alpha)$) in Fig.~\ref{FigScal} in order
to sort out the effect of field scaling. If the scaling approach works properly, we expect that $J_\mathrm{c}(B\epsilon)$ in the right panel has the same slope at all angles, which is indeed obtained for
$\gamma=3.5$. This value also seems realistic in view of the available data, i.e. $\gamma$ is 5-8 near
$T_\mathrm{c}$ and decreases with temperature.\cite{Kid11,Jar08,Jia08}

Although the field scaling brings the minima and maxima of
$J_\mathrm{c}$ close together, the curves do not collapse (left
panel in Fig.~\ref{FigScal}). The positions of the maxima do not
coincide, which could be caused by a higher anisotropy in the
unirradiated crystal. It also seems that the value of
$J_\mathrm{c}$ at the second maximum increases at large angles, but
this might be an artifact of the measurement method (VLF
currents). On the other hand, the decrease of the $J_\mathrm{c}$-minimum with $\alpha$ cannot be caused by VLF currents and
agrees with the overall behavior of the angular dependence of
$J_\mathrm{c}$ in the irradiated crystal.

Scaling of $J_\mathrm{c}$ in the irradiated crystals can be
performed by
$J_\mathrm{c}(B,\alpha)=A_J(\epsilon(\alpha))J_\mathrm{c}(B\epsilon(\alpha),0)$,
with an a priori unknown function $A_J(\alpha)$ (or $A_J(\epsilon(\alpha))$). (Note that this scaling fails at
very low fields, where the self field rotates $B$ within the sample
toward the $c$-axis. This is an artifact of our evaluation, because the applied self field correction only calculates the absolute value of $B$. Although correcting for $\alpha$ was possible, it would impede plotting
$J_\mathrm{c}(B,\alpha)$ without interpolation between measurements at different angles.) $A_J(\alpha)$ is expected to be
constant \cite{Bla92} within the single vortex pinning regime of collective pinning theory, where the defects are
assumed to be smaller than the coherence length $\xi$. In a simple
model, the pinning energy of a single defect becomes proportional
to $E_\mathrm{c}r_\mathrm{d}^3$, with the condensation energy
density $E_\mathrm{c}$ and the defect radius $r_\mathrm{d}$. Therefore,
the pinning energy does not depend on $\alpha$. If the defects are
larger than the coherence length, the pinning energy becomes
proportional to $E_\mathrm{p}\propto
E_\mathrm{c}r_\mathrm{d}\xi_{ab}\xi(\alpha)=E_\mathrm{c}r_\mathrm{d}\xi_{ab}^2\epsilon(\alpha)$,
thus decreases with $\alpha$. This qualitatively explains our data
on the irradiated crystal, although a direct scaling with pinning
energy $A_J(\alpha)\propto E_\mathrm{p}(\alpha)$ is not consistent
with our data, when assuming a realistic systematic error caused
by the VLF currents. However, a direct proportionality between
pinning energy and critical current density is not expected from
most pinning models, in particular in view of the changing elastic
properties of the vortex lattice when the field orientation
changes. Scaling by the square root of the pinning energy leads to
reasonable agreement of all data, but a quantitative analysis of
the angular dependence of $A_J$ is not meaningful because
of the systematic error of angular resolved magnetization
measurements. 

The angular dependence of $A_J$ should not be related to a particular superconductor, but should result from large pinning centers. Indeed, a decreasing $J_\mathrm{c}$ with increasing $\alpha$ was also observed in neutron irradiated coated conductors \cite{Eis10} before the intrinsic peak close to $H\| ab$ occurs and only if the field is significantly below $B_\mathrm{irr}$. 

The fishtail effect induces additional complexity into the angular
dependence of $J_\mathrm{c}$ (e.g. left panel of
Fig.~\ref{Figangular}, which can be understood by field scaling
(see above).) However, we find a crossover in $A_J(\alpha)$, which
decreases with $\alpha$ at low fields, but increases at higher
fields, in particular near the second peak. The behavior near and
above the second peak is essentially consistent with the
predictions of the anisotropic scaling approach,\cite{Bla92} if
one assumes the anisotropy to be a little higher and relates the
slightly different currents at the peak to the peculiarities
of the measurement method. At low fields on the other hand, the
currents decrease with $\alpha$, as in the irradiated crystal. The
crossover suggests that pinning in the pristine crystals is
dominated by comparatively large defects of low
density  at low fields and by small defects of high density at high magnetic
fields.

\section{Conclusions}
The angular dependence of the critical currents was derived from
magnetization measurements of Nd-1111 single crystals. The
fishtail effect and the introduction of disorder by neutron irradiation were shown to change the
current anisotropy significantly. It was demonstrated that the
originally proposed pure field scaling resulting from collective
pinning theory is valid only in a limited field range.
However, the concept can be extended to other pinning regimes by
introducing an additional $J_\mathrm{c}$-scaling, which was
motivated by the expected anisotropy of the pinning energy. This
extension is mandatory for $J_\mathrm{c}$ at low fields in the
unirradiated sample and in the whole field range after fast
neutron irradiation, since pinning is dominated by large defects
in both these cases.

\begin{acknowledgments}
We wish to thank H. W. Weber for fruitful discussions. This work was supported by the Austrian Science Fund (FWF): P22837-N20 and by the European-Japanese collaborative project SUPER-IRON (No. 283204).
\end{acknowledgments}


\end{document}